\documentclass[preprint,preprintnumbers,amsmath,amssymb,aps]{revtex4}

\usepackage{graphicx}       
\usepackage{dcolumn}        
\usepackage{bm}             
\usepackage{psfrag}

\newcommand{\nn}{\nonumber}
\newcommand{\grad}{\mbox{\boldmath $\nabla$}}

\newcommand{\gam}{\gamma}
\newcommand{\bet}{\beta}

\newcommand{\sig}{\sigma}

\newcommand{\brk}{\, & \,}
\newcommand{\lend}{\, \\ \,}

\newcommand{\tMor}{2M/r}

\begin{document}

\preprint{}

 \title{Instability of the massive Klein-Gordon field on the Kerr spacetime}

\author{Sam R. Dolan}
 \email{sam.dolan@ucd.ie}
 \affiliation{%
 School of Mathematical Sciences, University College Dublin, Belfield, Dublin 4, Ireland \\
}%

\date{\today}

\begin{abstract}
We investigate the instability of the massive scalar field in the vicinity of a rotating black hole. The instability arises from amplification caused by the classical superradiance effect. The instability affects bound states: solutions to the massive Klein-Gordon equation which tend to zero at infinity. We calculate the spectrum of bound state frequencies on the Kerr background using a continued fraction method, adapted from studies of quasinormal modes. We demonstrate that the instability is most significant for the $l = 1$, $m = 1$ state, for $M \mu \lesssim 0.5$. For a fast rotating hole ($a = 0.99$) we find a maximum growth rate of $\tau^{-1} \approx 1.5 \times 10^{-7} (GM/c^3)^{-1}$, at $M \mu \approx 0.42$. The physical implications are discussed.
\end{abstract}

\maketitle

%
%

\section{\label{sec:introduction}Introduction}

The issue of black hole stability was first addressed over fifty years ago. In an influential study, Regge and Wheeler \cite{Regge-1957} showed that the Schwarzschild solution is stable. If a Schwarzschild black hole is perturbed slightly, then the perturbation will oscillate and die away, rather than grow, over time \cite{Vishveshwara-1970}. Some fraction of the initial perturbation is absorbed through the event horizon, and the remainder is radiated away to infinity. The demonstration of stability contributed to a growing belief that black holes could be more than mere mathematical curiosities. 

When Kerr \cite{Kerr-1963} published a new black hole solution, describing the gravitational field induced by a rotating point mass, the stability question was addressed anew. Press and Teukolsky \cite{Teukolsky-1972, Press-1973} showed that the Kerr solution remains stable under gravitational perturbations. However, the situation is complicated somewhat by an effect known as \emph{superradiance}. Certain perturbations are enhanced by the rotation of the hole. The energy radiated away to infinity may actually exceed the energy present in the initial perturbation. In effect, perturbations may extract rotational energy from the hole. Since superradiance affects classical fields, it cannot be dismissed as a purely quantum phenomenon. 

Various authors have shown that, for superradiance to occur, the oscillation frequency of the perturbation $\omega$ must be less than a critical value $\omega_c$, given by
\begin{equation}
\omega_c = \frac{a m}{2 M r_+} . \label{superr}  
\end{equation}
Here, $M$ is the black hole mass, $a = J / M$ is the rotation rate of the hole, $m$ is the azimuthal number of the perturbation, and $r_+$ is the radius of the outer event horizon.

Press and Teukolsky \cite{Press-1972} postulated that, if the superradiance emerging from a perturbed hole were reflected back onto the hole, then an initially small perturbation could be made to grow without bound. This is the so-called ``black hole bomb'' idea \cite{Cardoso-2004-bomb}. To reflect the radiation, Press and Teukolsky suggested using a special arrangement of mirrors. However, reflection will also occur naturally if the perturbing field has a rest mass \cite{Damour-1976, Zouros-1979, Detweiler-1980, Furuhashi-2004, Cardoso-2005-superr}. In this case, a spectrum of `bound states' is present \cite{Lasenby-2005-bs} (also referred to in the literature as `resonances' \cite{Damour-1976} or `quasi-stationary levels' \cite{Gaina-1993}). Bound states are localised in the black hole potential well and tend to zero at spatial infinity. Inevitably, bound states have complex frequencies, as flux passes one way through the (outer) event horizon. The imaginary part of the frequency determines the rate at which the perturbation decays (or grows) with time. The bound states idea has been explored by a number of authors \cite{Deruelle-1974, Damour-1976, Ternov-1978, Ternov-1980, Gal'tsov-1983, Gaina-1992, Gaina-1993, Lasenby-2005-bs, Grain-2007, Laptev-2006} over the years.  

The nature of the bound state spectrum and the superradiant instability depend on two parameters. The first is the rotation speed of the hole $a$. The second is the (dimensionless) product of the black hole mass $M$ and the field mass $\mu$. The product $M \mu$ is equivalent to the ratio of the event horizon size to the perturbing field's Compton wavelength,
\begin{equation}
M \mu \equiv \frac{G M \mu}{\hbar c} \sim \frac{r_h}{\lambda_C} .
\end{equation} 

Some years ago, Detweiler \cite{Detweiler-1980} studied the growth rate of  bound states in the limit $M \mu \ll 1$. He estimated an e-folding time $\tau$ for the first co-rotating state ($l = 1$, $m = 1$) to be
\begin{equation}
\tau \, \approx \, 24 (a/M)^{-1} (\mu M)^{-9} \, (GM / c^3) .
\label{eq-detweiler}
\end{equation}
Zouros and Eardley \cite{Zouros-1979} computed an approximation for the growth rate in the opposite limit, $M \mu \gg 1$, using the JWKB approximation. They found
\begin{equation}
\tau \approx 10^7 e^{1.84 M \mu}  (GM / c^3) .
\label{eq-zouros}
\end{equation} 
Given these results, when might the instability be significant? For a pion around a solar-mass black hole, $M \mu \sim 10^{18}$. Clearly, from (\ref{eq-zouros}), the instability is insignificant for astrophysical black holes, unless there exists an unknown particle with a tiny but non-zero rest mass. However, the instability may be important for primordial black holes \cite{Zouros-1979}. Thus, an investigation of the instability in the regime $M\mu \sim 1$ is well-motivated. A numerical study of the Kerr-Newman instability in this regime was recently conducted \cite{Furuhashi-2004}. In this paper we develop alternative methods to study the uncharged case.

This paper has four broad aims. First, to show the existence of a spectrum of bound states of the massive Klein-Gordon field on the Kerr background. Second, to show that these bound states decay with time if $\text{Re}(\omega) < \omega_c$, but grow with time if $\text{Re}(\omega) > \omega_c$. Third, to show that bound state frequency spectra may be calculated numerically using a simple continued-fraction method \cite{Leaver-1985}, previously used to compute quasinormal modes (QNMs). Fourth, to compute accurate upper limits on the bound state growth rates, and estimate the physical implications. 

The remainder of the paper is organised as follows. Section \ref{sec-analytic} examines the analytic properties of the massive Klein-Gordon field on the Kerr background. To begin,  we introduce two alternative coordinate systems, and separate the field into radial and angular parts (\ref{sec-coord-sys}). Next, by considering the flux of stress-energy across the outer horizon, we derive a simple expression for the time-evolution of the field (\ref{sec-stress-energy}). Then, we set up the wave equation on the Kerr background (\ref{sec-equations}) and define boundary conditions for bound state modes (\ref{sec-boundstates}). Section \ref{sec-ctdfrac} outlines the continued-fraction method \cite{Leaver-1985} which we employ to compute bound state spectra numerically. In section \ref{sec-results} we present the results of this numerical approach. Bound state spectra are presented in \ref{sec-bs-schw} and \ref{sec-bs-kerr}, and the Kerr instability is examined in detail in \ref{sec-instability}. We conclude in section \ref{sec-discussion} by discussing the physical relevance of our results, and some prospects for further work.

Note that throughout this paper we adopt the spacetime signature $[1,-1,-1,-1]$, and natural units, $G = \hbar = c = 1$.

\section{The Klein-Gordon field on the Kerr background\label{sec-analytic}}
In this section we examine the (classical) Klein-Gordon field $\Phi(x^\mu)$ on the Kerr black hole background. First, we introduce two alternative coordinate systems to describe the Kerr spacetime, and discuss the separation of variables in these systems. Next, we look at the field's stress-energy tensor, which satisfies a simple conservation law. By applying Gauss' theorem, we show that the time-evolution of the field is directly related to a superradiance condition \cite{Zouros-1979}. Then, we formulate the wave equation on the Kerr background, and examine the asymptotic behaviour of the field as $r \rightarrow r_+$ and $r \rightarrow \infty$. Finally, we define boundary conditions for bound states, and show that in the non-relativistic limit, the spectrum of frequencies is hydrogenic \cite{Lasenby-2005-bs}.
 
\subsection{Coordinate systems\label{sec-coord-sys}}
The Kerr spacetime is described by a number of coordinate systems. The most commonly used are Boyer-Lindquist coordinates, $x^\mu = [t,r,\theta,\phi], \; \mu = 0 \dots 3$, for which the line element takes the form
\begin{align}
ds^2 &= \left(1 - \frac{2Mr}{\rho^2} \right) dt^2 + \frac{4aMr \sin^2 \theta}{\rho^2} dt d\phi - \frac{\rho^2}{\Delta} dr^2 - \rho^2 d\theta^2 \nn \\ 
 & \quad \quad - \left[(r^2 + a^2) \sin^2 \theta + \frac{2Mr}{\rho^2} a^2 \sin^4 \theta \right] d\phi^2   \label{BLmetric}
\end{align}
where 
\begin{equation}
\Delta = r^2 - 2Mr + a^2  \quad \text{ and } \quad \rho^2 = r^2 + a^2 \cos^2 \theta . 
\end{equation}
The Kerr solution has two event horizons, at $r_\pm = M \pm \sqrt{M^2 - a^2}$, and two stationary limit surfaces, at $r_{S\pm} = M \pm \sqrt{M^2 - a^2 \cos^2 \theta}$. An advantage of Boyer-Lindquist coordinates is that the metric has only one off-diagonal term, $dt d\phi$, whereas a disadvantage is that it takes an infinite coordinate time $t$ for ingoing geodesics to cross the outer horizon at $r = r_+$.   

Perhaps of more interest from a physical point of view are ingoing Kerr coordinates, $\tilde{x}^\mu = [\tilde{t}, r, \theta, \tilde{\phi}]$, in which the contravariant metric tensor $\tilde{g}^{\mu \nu}$ takes the form 
\begin{equation}
\tilde{g}^{\mu \nu} = \frac{1}{\rho^2} \begin{pmatrix} \rho^2 + 2Mr & -2Mr & 0 &0 \\ -2Mr & -\Delta & 0 & -a \\ 0 & 0 & -1 & 0 \\ 0 & -a & 0 & -\tfrac{1}{\sin^2 \theta} \end{pmatrix}
\end{equation}
In this coordinate system, the ingoing principal null geodesics are straight lines, $dr = -d\tilde{t}$, and ingoing geodesics pass through the outer horizon in a finite coordinate time.

The $r$ and $\theta$ coordinates are the identical in both coordinate systems, and $\tilde{t}$ and $\tilde{\phi}$ are related to $t$ and $\phi$ by
\begin{equation}
\tilde{t} = t + \alpha(r), \quad \quad \tilde{\phi} = \phi + \beta(r),
\end{equation}
where
\begin{align}
\alpha(r) &= -r + \int \frac{r^2 + a^2}{\Delta} dr = \frac{2r_+}{r_+ - r_-} \ln | r - r_+ | - \frac{2 r_-}{r_+ - r_-} \ln | r - r_- | ,  \\
\beta(r)  &= \int \frac{a}{\Delta} dr = \frac{a}{r_+ - r_-} \ln | r - r_+ | - \frac{a}{r_+ - r_-} \ln | r - r_- | .
\end{align}
Following \cite{Brill-1972}, we assume that the Klein-Gordon field $\Phi(x^\mu)$ may be separated into a product of one-dimensional functions,
\begin{equation}
\Phi(x^\mu) =  e^{i m \phi} e^{-i \omega t} S_{lm}(\theta) R_{lm}(r) . \label{sep-var-1}
\end{equation}
The frequency $\omega$ is permitted to be complex. The sign of $\text{Im}(\omega)$ determines whether the solution is decaying ($\text{Im} (\omega) < 0$) or growing ($\text{Im} (\omega) > 0$) in time.

Since $\Phi$ is a scalar, it is unchanged by a (passive) coordinate transformation. However, the separation of variables is modified, so that
\begin{equation}
\Phi(x^\mu) = \tilde{\Phi}(\tilde{x}^\mu) = e^{-i \omega t} e^{i m \phi} S_{lm}(\theta) R_{lm}(r) = e^{-i \omega \tilde{t}} e^{i m \tilde{\phi}} S_{lm}(\theta) \tilde{R}_{lm} (r) . \label{sep-var-2}
\end{equation}
Thus, the radial functions are related by
\begin{equation}
\tilde{R}_{lm}(r) = e^{i \omega \alpha(r)} e^{-i m \beta(r)} R_{lm}(r). \label{Rtilde}
\end{equation}

\subsection{Stress-Energy Conservation and Time-Evolution\label{sec-stress-energy}}
Here we seek to relate the time-evolution of the field to the superradiance condition. To do so, we integrate the stress-energy of the field $\Phi$ over a four-volume with its boundary at the (outer) event horizon. A similar argument was outlined in \cite{Zouros-1979}.

The Lagrangian density $\mathcal{L}$ of a minimally-coupled (complex) scalar field is
\begin{equation}
\mathcal{L} = \tfrac{1}{2} g^{\mu \nu} \partial_{(\mu} \Phi^\ast \, \partial_{\nu)} \Phi - \tfrac{1}{2} \mu^2 |\Phi|^2
\end{equation}
where $ \partial_{(\mu} \Phi^\ast \partial_{\nu)} \Phi = \tfrac{1}{2}(\partial_\mu \Phi^\ast \partial_\nu \Phi + \partial_\nu \Phi^\ast \partial_\mu \Phi)$. The symmetric stress-energy tensor ${T^\mu}_\nu$ is
\begin{equation}
T^\mu_\nu = g^{\mu \lambda} \partial_{(\lambda} \Phi^\ast \, \partial_{\nu)} \Phi - \delta^\mu_\nu \mathcal{L}
\end{equation}
The stress-energy tensor satisfies the conservation law ${T^\mu}_{\nu ; \mu} = 0$. Since the Kerr spacetime is stationary, there exists a Killing vector $\xi^\mu$ conjugate to the time coordinate with components $\xi^\mu = [1, 0, 0, 0]$.  The Killing vector is timelike outside the stationary limit surface at $r = r_{S+}$, but spacelike within. Contracting the stress-energy tensor with this Killing vector and using Killing's equation $\xi_{\mu ; \nu} = \xi_{\nu ; \mu}$ yields
\begin{equation}
\left( {T^\mu}_\nu \xi^\nu \right)_{; \mu} = \left( {T^\mu}_0 \right)_{; \mu}   = 0. \label{stress-energy-cons}
\end{equation}

We now seek to relate the flux of stress-energy crossing the horizon to the time-evolution of the field. For this, we use the ingoing-Kerr coordinate system. Let us construct a semi-infinite four-volume: an infinitessimal time-slice of width $\Delta \tilde{t}$ exterior to the outer event horizon. This is bounded by upper and lower hypersurfaces ($\tilde{t} = \pm \Delta \tilde{t} / 2$ and $r > r_+$), and a null hypersurface at the horizon ($r = r_+$). Applying Gauss's theorem to (\ref{stress-energy-cons}), and taking the limit $\Delta t \rightarrow 0$, leads to
\begin{equation}
\int_{\partial \Sigma} {T^\mu}_0 k_\mu d\Omega  = - \frac{\partial}{\partial \tilde{t}} \left( \int_{\Sigma} {T^\mu}_0 n_\mu d^3 x  \right)  \label{Tcons}
\end{equation}
where the surfaces $\partial \Sigma$ and $\Sigma$, and normal vectors $n_\mu$ and $k_\mu$ are defined by
\begin{align}
\partial \Sigma : \quad & \tilde{t} = 0, \; r = r_+, & k_\mu &= -\delta^1_\mu, & d\Omega &= \rho^2 \sin \theta d\theta d\tilde{\phi} , \\
\Sigma : \quad & \tilde{t} = 0, \; r > r_+, & n_\mu &= (\tilde{g}^{00})^{-1/2} \, \delta^0_\mu, & d^3x &= (\tilde{g}^{00})^{1/2} \rho^2 \sin \theta dr d\theta d\tilde{\phi} .
\end{align}

To proceed further, we introduce the separation of variables (\ref{sep-var-1}). The time-derivative results in an overall factor of $2 \, \text{Im}(\omega)$ on the right-hand side of (\ref{Tcons}). The integral over the horizon is straightforward to compute:  
\begin{align}
\int_{\partial \Sigma} {T^{\mu}}_0 n_\mu d\Omega &= - \rho^2 \int_{0}^\pi \int_{0}^{2\pi} {T^{1}}_{0} \sin \theta d\theta d\tilde{\phi} \nn \\
 &= \left( 2 M r_+ |\omega|^2 - a m \text{Re}(\omega) \right) |\tilde{R}_{lm}(r_+)|^2 . \label{se-lhs}
\end{align}
Here we have assumed the angular functions are normalised so that $\int |S_{lm}(\theta)|^2 \sin \theta d\theta = 1/2\pi$. Hence, the imaginary part of $\omega$ is 
\begin{equation}
2 \, \text{Im} ({\omega}) = -\left( 2 M r_+ |\omega|^2 - a m \text{Re}(\omega) \right)  \frac{|\tilde{R}_{lm}(r_+)|^2}{\int_\Sigma {T^0}_0 d^3 x  } .   \label{Im-omega}
\end{equation}
Therefore, unstable (exponentially-growing) states are possible if 
\begin{equation}
\frac{|\omega|^2}{\text{Re}(\omega)} < \frac{a m }{2Mr_+} . \label{superr2} 
\end{equation}
If $|\text{Re}(\omega)| \gg |\text{Im}(\omega)|$, equation (\ref{superr2}) reduces to the superradiance condition (\ref{superr}).

The validity of this argument rests on three conditions. First, that the radial function $\tilde{R}(r_+)$ is finite at the horizon. Second, that the radial solution is normalisable over $r > r_+$. Third, that the integral $\int_\Sigma {T^0}_0 d^3 x $ in the denominator of (\ref{Im-omega}) is non-zero and positive. In section \ref{sec-boundstates}, we show that the boundary conditions imposed on the bound states ensure that the first two conditions are met. The third condition requires further analysis. In the ingoing-Kerr frame, the ``energy'' component of the stress-energy is 
\begin{equation}
{T^0}_0 = \frac{1}{2 \rho^2} \left[ \left((\rho^2 + 2Mr)|\omega|^2 + \rho^2 \mu^2 \right)|\Phi|^2 + \Delta |\partial_r \Phi|^2 + |\partial_\theta \Phi|^2 +  \frac{1}{\sin^2 \theta} |\partial_\phi \Phi|^2 + 2 a m J_r \right],
\end{equation}
where $J_r$ is a radial current given by
\begin{equation}
J_r = \frac{1}{2i} \left( \Phi^\ast \partial_r \Phi - \Phi \partial_r \Phi^\ast \right) .
\end{equation}
${T^0}_0$ is positive everywhere outside $r = r_{S+}$. However, ${T^0}_0$ can be negative inside the ergosphere (where the Killing vector is spacelike) if there is sufficient coupling between radial and azimuthal currents. If $\int_\Sigma {T^0}_0 d^3 x$ were to pass through zero then the growth rate given by (\ref{Im-omega}) would tend to infinity! However, this is not observed in the numerical results of section \ref{sec-results}.  

\subsection{The Wave Equation \label{sec-equations}}
In this section we separate the Klein-Gordon field equation in Boyer-Lindquist coordinates on the Kerr background \cite{Brill-1972}, and examine its asymptotic behaviour as $r \rightarrow r_+$ and $r \rightarrow \infty$. 

The wave equation for a scalar field $\Phi$ of mass $\mu$ is 
\begin{equation}
\Box \Phi + \mu^2 \Phi = (-g)^{-1/2} \partial_\mu \left( (-g)^{1/2} g^{\mu \nu} \partial_\nu \Phi \right) + \mu^2 \Phi = 0
\end{equation}
where $g = \det(g_{\mu \nu})$. Employing the Boyer-Lindquist metric (\ref{BLmetric}), 
\begin{align}
\left( \frac{(r^2 + a^2)^2}{\Delta} - a^2 \sin^2 \theta \right) \partial_t \partial_t \Phi  + \frac{4mar}{\Delta} \partial_t \partial_\phi \Phi + \left( \frac{a^2}{\Delta} - \frac{1}{\sin^2 \theta} \right) \partial_\phi \partial_\phi \Phi  & \\
- \partial_r \left( \Delta \, \partial_r \Phi \right) - \frac{1}{\sin \theta} \, \partial_\theta \left( \sin \theta \, \partial_\theta \Phi\right) + \mu^2 \rho^2 \Phi &= 0 .
\end{align}
Decomposing the field with the ansatz (\ref{sep-var-1}) leads to ordinary differential equations for the radial function 
\begin{equation}
\frac{d}{dr} \left( \Delta \frac{d R_{lm}}{d r} \right) + \left[ \frac{\omega^2 (r^2 + a^2)^2 - 4Mam\omega r + m^2 a^2 }{\Delta} - \left( \omega^2 a^2 + \mu^2 r^2 + \Lambda_{lm} \right) \right] R_{lm}(r) = 0 , \label{radialeq}
\end{equation}
and the angular function
\begin{equation}
\frac{1}{\sin \theta} \frac{d}{d \theta} \left(\sin \theta \frac{d S_{lm}}{d \theta} \right) + \left[ a^2 (\omega^2 - \mu^2) \cos^2 \theta - \frac{m^2}{\sin^2 \theta} + \Lambda_{lm} \right] S_{lm}(\theta)  = 0 .
\end{equation}
From here on, we often set $M = 1$, so that $r$ and $a$ are measured in units of $M$, and $\omega$ and $\mu$ in units of $M^{-1}$. 

The angular solutions are spheroidal harmonics \cite{Abramowitz-1972, Berti-2005},  $S_{lm} = S_l^m(\cos \theta; c)$. The degree of spheroidicity depends on the parameter $c = a \sqrt{\omega^2 - \mu^2}$. In the non-rotating limit, the spheroidal harmonics reduce to spherical harmonics, $S_l^m \rightarrow Y_l^m$, and $\Lambda_{lm} \rightarrow l (l +1)$. Numerical methods for computing the angular separation constant $\Lambda_{lm}$ are discussed in section \ref{sec-ctdfrac}.

Close to the outer horizon the radial solutions go as
\begin{equation}
\lim_{r \rightarrow r_+} R_{lm}(r) \sim (r - r_+)^{\pm i \sigma}, \quad \quad \text{ where } \quad \sigma = \frac{2 r_+ (\omega - \omega_c)}{r_+ - r_-}  . \label{asympt-rplus}
\end{equation}
The choice of sign in the exponent determines the behaviour at the horizon. The negative sign is the correct choice for an ingoing wave (as measured by a comoving observer). 

Towards spatial infinity, the radial function has the asymptotic behaviour
\begin{equation}
\lim_{r \rightarrow \infty} R_{lm}(r) \sim r^{-1} \, r^{(\mu^2 - 2\omega^2)/ q} \, \exp(q r), \quad \quad \text{ where } \quad q = \pm \sqrt{\mu^2 - \omega^2} . 
\end{equation}
The sign of the real part of $q$ determines the behaviour of the wavefunction as $r \rightarrow \infty$. If $\text{Re}(q) > 0$, the solution diverges, whereas if $\text{Re}(q) < 0$, the solution tends to zero. The general solution is a linear sum of solutions with both types of behaviour, so is also divergent. 

\subsection{Bound States\label{sec-boundstates}}
By definition, bound state solutions are ingoing at the horizon, and tend to zero at infinity ($\text{Re}(q) < 0$). On the other hand, the well-known quasinormal modes \cite{Chandrasekhar-1983} are ingoing at the horizon and purely outgoing (and divergent) at infinity ($\text{Re}(q) > 0$). In both cases, imposing a pair of boundary conditions leads to a discrete spectrum of (complex) frequencies.

One feature of the ingoing boundary condition ($R \rightarrow (r-r_+)^{-i \sigma}$ as $r \rightarrow r_+$) is that decaying states with $\text{Im} (\omega) < 0$ have a radial function $R(r)$ which is divergent at the horizon. This is due to the fact that Boyer-Lindquist coordinates are only valid in the exterior region, $r > r_+$. They fail to describe classical geodesics that cross the outer horizon: it takes an infinite coordinate time $t$ to cross from $r > r_+$ to $r < r_+$. 

To study states that pass through the horizon, and to interpret the physical content of the wavefunction, we may use ingoing-Kerr coordinates. By applying (\ref{Rtilde}) to (\ref{asympt-rplus}), it is straightforward to verify that the ingoing radial function $\tilde{R}_{lm}(r)$ is regular at the outer horizon. However, the outgoing solution remains divergent. Physically, this implies that an infalling observer measures the local probability density of an ingoing solution to be finite and well-defined. On the other hand, the probability density of the outgoing solution is undefined at $r = r_+$. 

As mentioned above, a pair of boundary conditions gives rise to a discrete spectrum of frequencies. The spectrum of the (classical) Dirac equation on the Schwarzschild background has been studied in depth by Lasenby \emph{et al.} \cite{Lasenby-2005-bs}. In the limit $M \mu \ll l$, the bound state spectrum of the massive Dirac field  resembles that of the hydrogen atom. That is,
\begin{equation}
\hbar \omega_n \approx \left( 1 - \frac{M^2 \mu^2}{2 \bar{n}^2} \right)  \mu c^2 , \label{hydrogen-spectrum}
\end{equation}
where $\bar{n} = n + l + 1$ is the principal quantum number of the state. In Appendix \ref{non-rel} we show that the scalar field also has a hydrogenic spectrum  in this limit. To lowest order in $M\mu$, the spectrum is not dependent on the rotation of the hole or the spin of the field.  

\section{Numerical Method\label{sec-ctdfrac}}
In this section, we show that both the quasinormal modes (QNMs) \emph{and} the bound states of the massive scalar field can be found by numerically solving a three-term recurrence relation. In a classic study, Leaver \cite{Leaver-1985} was the first to use this approach to find the QNMs of the gravitational field. More recently, Konoplya and Zhidenko \cite{Konoplya-2006} applied the same method to find the QNMs of the massive scalar field. Cardoso and Yoshida \cite{Cardoso-2005-superr} have shown that massive scalar bound states can be found by solving a five-term recurrence relation. Here, we instead derive a three-term relation and apply the continued-fraction method.

The radial equation (\ref{radialeq}) has received much attention. Equations of this form first arose in a study of the electronic spectrum of the hydrogen molecule \cite{Jaffe-1934, Baber-1935}, over seventy years ago. Equation (\ref{radialeq}) may be transformed to a singly-confluent Heun equation \cite{Fiziev-2006} by a suitable substitution. Heun equations have four regular singular points. In the singly-confluent case, two of these points are merged together at $r = \infty$. The other singular points lie at the horizons, $r=r_+$ and $r=r_-$.  The singly-confluent equation is related to the ``generalised spheroidal equation'' \cite{Leaver-1986}. In the critically-rotating case ($a = 1$), both horizons are combined at $r = M$, and (\ref{radialeq}) is related to the doubly-confluent Heun equation.

In section \ref{sec-boundstates}, we discussed appropriate boundary conditions at $r = r_+$ and $r = \infty$. With this in mind, let us now look for a solution of the form 
\begin{equation}
R(r) = (r - r_+)^{-i \sig} (r - r_-)^{i \sig + \chi - 1} e^{q r} \sum_{n=0}^\infty a_n \left(\frac{r - r_+}{r - r_-}\right)^n   \label{Rsubs}
\end{equation}
where 
\begin{equation}
\sigma = \frac{2 r_+ (\omega - \omega_c)}{r_+ - r_-}, \quad \quad 
q = \pm \sqrt{ \mu^2 - \omega^2 }, \quad \quad \text{ and } \quad \quad \chi = \frac{\mu^2 - 2 \omega^2}{q} .
\end{equation}
The choice of the sign of the real part of $q$ determines the behaviour of the wavefunction as $r \rightarrow \infty$. If $\text{Re}(q) > 0$, the solution diverges towards infinity, whereas if $\text{Re}(q) < 0$ the solution tends to zero. Therefore, the same method can be applied to look for both quasinormal modes (by choosing $\text{Re}(q) > 0$) \emph{and} the bound state modes (by choosing $\text{Re}(q) < 0$). 

Substituting (\ref{Rsubs}) into the radial equation (\ref{radialeq}) yields a three-term recurrence relation for the coefficients $a_n$. Adopting Leaver's nomenclature, we find
\begin{align}
\alpha_0 a_1 + \beta_0 a_0 &= 0 \\
\alpha_n a_{n+1} + \beta_n a_n + \gamma_n a_{n-1} &= 0,  \quad \quad n > 0, \quad n \in \mathbb{N},
\end{align}
where
\begin{align}
\alpha_n &= n^2 + (c_0 + 1) n + c_0 ,  \\
\beta_n   &= -2n^2 + (c_1 + 2)n + c_3 , \\
\gamma_n &= n^2 + (c_2 - 3)n + c_4 .
\end{align}
The constants $c_0$, $c_1$, $c_2$, $c_3$ and $c_4$ are somewhat more complicated than in the massless case \cite{Leaver-1985}. Explicitly, 
\begin{align}
c_0 &= 1 - 2i\omega - \frac{2 i}{b} \left( \omega - \frac{am}{2} \right), \label{eqc0} \\
c_1 &= -4 + 4i(\omega - iq(1 + b)) + \frac{4i}{b} \left( \omega - \frac{am}{2} \right) - \frac{2 (\omega^2 + q^2)}{q} , \label{eqc1} \\
c_2 &= 3 - 2i\omega - \frac{2(q^2 - \omega^2)}{q} - \frac{2i}{b} \left( \omega - \frac{a m}{2} \right) , \label{eqc2}  \\
c_3 &= \frac{2i(\omega - iq)^3}{q} + 2 (\omega - iq)^2 b + q^2 a^2 + 2iqam - \Lambda_{lm} - 1 - \frac{(\omega - iq)^2}{q} + 2qb \nonumber \\
        & \quad \quad + \frac{2i}{b} \left( \frac{(\omega - iq)^2}{q} + 1 \right) \left( \omega - \frac{a m}{2} \right) , \label{eqc3} \\
c_4 &= \frac{(\omega - iq)^4}{q^2} + \frac{2i \omega (\omega - iq)^2}{q} - \frac{2i}{b} \frac{(\omega - iq)^2}{q} \left(\omega - \frac{am}{2}\right) . \label{eqc4}
\end{align}
where
\begin{equation}
b = \sqrt{1 - a^2} .
\end{equation}
In the massless limit ($\mu = 0$, $q = i \omega$), equations (\ref{eqc0})---(\ref{eqc4}) reduce to Leaver's expressions \cite{Leaver-1985}.

The angular eigenvalue $\Lambda_{lm}$ (appearing in the expression for $c_3$) may be expanded as a power series 
\begin{equation}
\Lambda_{l m} = l (l + 1) + \sum_{k=1}^\infty f_{k} \, c^{2k}, \quad \text{where} \quad c^2 = a^2 ( \omega^2 - \mu^2 )
\end{equation}
Seidel \cite{Seidel-1989} lists the series expansion coefficients up to $f_4$. The power series provides an adequate approximation for $\Lambda_{lm}$ up to $c \sim l$. Beyond this regime, other methods are available. These include Leaver's continued-fraction method \cite{Leaver-1985} or Hughes' \cite{Hughes-2000} spectral decomposition method.



The ratio of successive coefficients $a_n$ is given by an infinite continued fraction
\begin{equation}
\frac{a_{n+1}}{a_n} = - \frac{\gam_{n+1}}{\beta_{n+1}-}\frac{\alpha_{n+1}\gam_{n+2}}{\beta_{n+2}-}\frac{\alpha_{n+2}\gam_{n+3}}{\beta_{n+3}-} \ldots
\end{equation}
Substituting $n = 0$ into the above expression and comparing with $a_1 / a_0 = - \beta_0 / \alpha_0$ leads to the implicit condition
\begin{equation}ß
\beta_0 - \frac{\alpha_0 \gam_1}{\bet_1 -} \frac{\alpha_1 \gamma_2}{\beta_2 -} \frac{\alpha_2 \gamma_3}{\beta_3 - } \ldots = 0.
\end{equation}
This condition is only satisfied for particular values of $\omega$ corresponding to the bound state (or QNM) frequencies. To find these values we employed a simple 2D minimisation algorithm.

\section{Results\label{sec-results}}
In this section we present frequency spectra determined numerically via the continued-fraction approach. First, to validate the code, we compute some QNM frequencies and compare with values in the literature. Next, we present bound state spectrum in the non-rotating and rotating cases. Finally, we verify the existence of unstable states (with $\text{Im}(\omega) > 0$) and compute their growth rates.

\subsection{Quasinormal Mode Spectra\label{sec-qnm}}
The continued-fraction method was first developed to find QNM frequencies \cite{Leaver-1985}. QNMs have been extensively studied through a variety of methods \cite{Leaver-1985, Andersson-1992, Simone-1992}. Hence, we begin by validating our continued-fraction code by computing QNM frequencies. 


Recently, Konoplya and Zhidenko \cite{Konoplya-2006} conducted a survey of the massive scalar QNMs using the continued fraction method. Their paper includes tables of QNM frequencies quoted to six decimal places, which we used to validate our numerical code. We found agreement to six decimal places with the frequencies listed in Tables I to V in \cite{Konoplya-2006}, apart from a few anomalies (probably introduced at the typesetting stage). In Appendix \ref{sec-tables} we include tables of QNM frequencies for the $n = 0$, $\,l = 1$, $\,m = -1$ mode (Table \ref{table-qnm1}) and the $n = 0$, $\,l = 2$, $\,m = 2$ mode (Table \ref{table-qnm2}). These tables are intended to help with any future numerical validation. 

\begin{figure}
\begin{center}
\includegraphics[width=12cm]{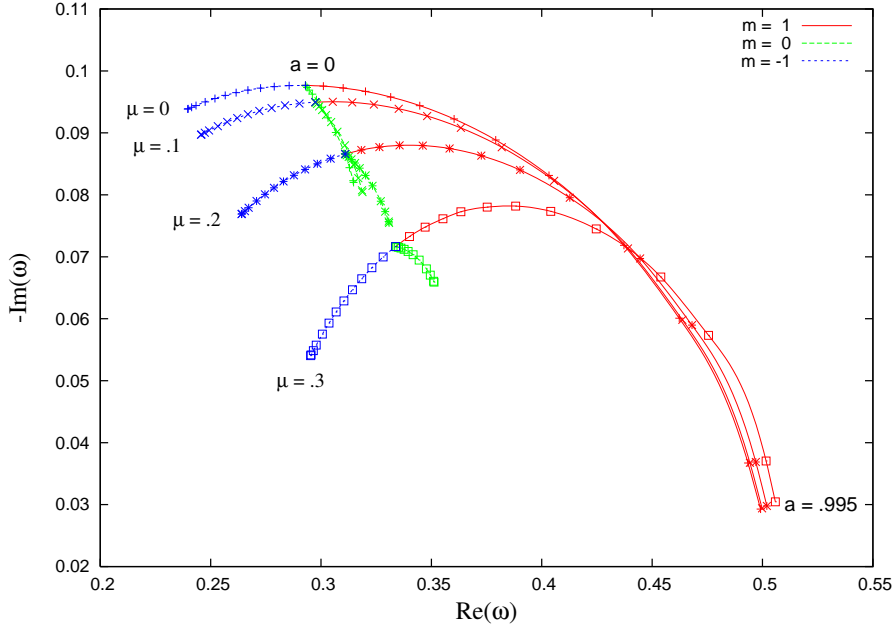}
\end{center}
\caption[]{\emph{The frequencies of the lowest $l = 1$ quasinormal modes}. The plot shows the QNM frequency as a function of black hole rotation $a$, for a variety of field masses, $\mu = 0$, $0.1$, $0.2$ and $0.3$. The $m = 1$ (right), $m = 0$ (middle) and $m = -1$ (left) branches are shown. The points shown are for the values $a = 0$, $0.1$, $0.2$, $0.3$, $0.4$, $0.5$, $0.6$, $0.7$, $0.8$, $0.9$, $0.95$, $0.99$ and $0.995$.    }
\label{fig-l1-qnm}
\end{figure}

Figure \ref{fig-l1-qnm} shows how the complex frequencies of the lowest $l = 1$, $m=-1 \ldots 1$ states vary with the field mass $\mu$ and black hole rotation speed $a$. The red (right) lines are for the co-rotating state, $m = 1$, the green (middle) lines are for $m=0$, and the blue (left) lines are for $m = -1$. In the non-rotating case $a = 0$, the lines meet, since in this case the frequency does not depend on $m$. 

The four sets of lines in Fig. \ref{fig-l1-qnm} correspond to the field masses $\mu = 0$, $0.1$, $0.2$ and $0.3$. As shown by Simone and Will \cite{Simone-1992}, larger field mass $\mu$ increases the oscillation frequency $\text{Re}(\omega)$ but decreases the damping $\text{Im}(\omega)$. Rotation $a$ also decreases the damping, particularly of the co-rotating state. Faster rotation increases the oscillation frequency of the $m =1$ state, but decreases the oscillation frequency of the $m = -1$ state.


\subsection{Bound State Frequencies: Schwarzschild $(a = 0)$\label{sec-bs-schw}}
From the analysis of section \ref{sec-stress-energy}, we expect all bound states on the Schwarzschild background to decay with time, since $\text{Im}(\omega) < 0$ by equation (\ref{Im-omega}). Furthermore, in the non-relativistic regime $M \mu \ll l$, we expect the frequencies to be approximated by (\ref{hydrogen-spectrum}). Our numerical results conform to both expectations. 

\begin{figure}
\begin{center}
\includegraphics[width=16cm]{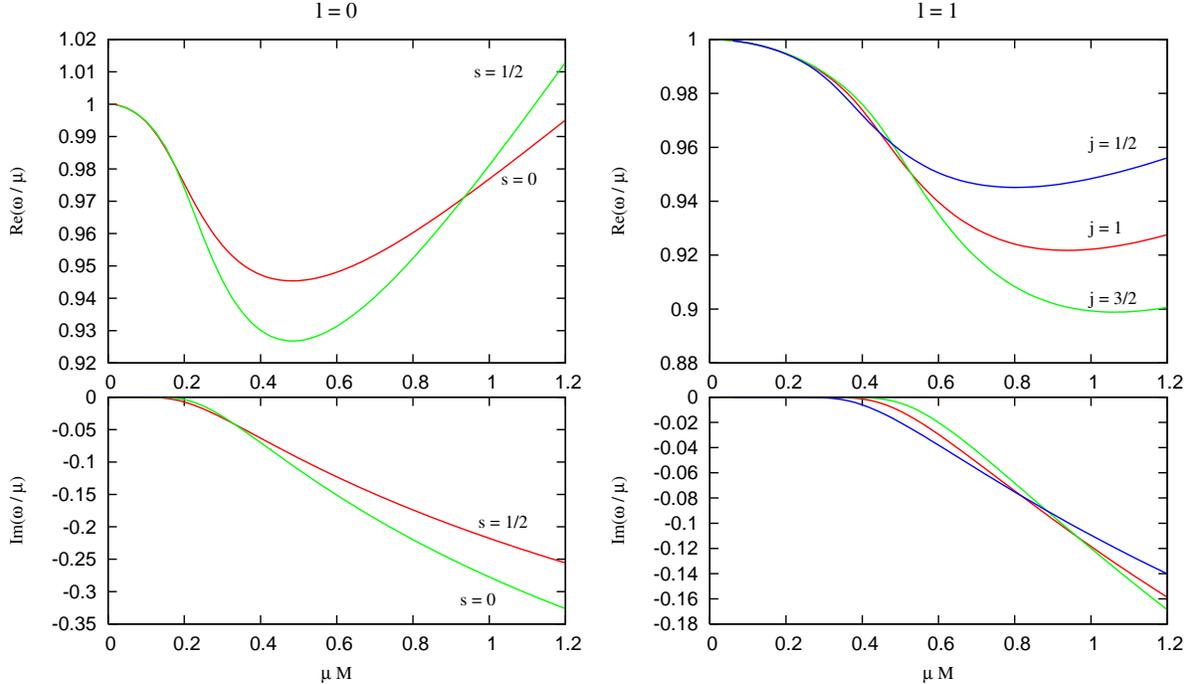}
\end{center}
\caption[]{\emph{Massive scalar $(s=0)$ and spinor $(s=1/2)$ bound state frequencies of the Schwarzschild hole}. The upper plots show the real component of energy (i.e. the oscillation frequency), and the bottom plots show the imaginary component (i.e. the decay frequency), as a function of gravitational coupling $M\mu$. The left plots compare the $l =0$ scalar ground state with the $j = 1/2$ spinor ground state. The right plots compare the $l = 1$ (scalar)  and the $j = 1/2$ and $j = 3/2$ (spinor) levels.}
\label{fig-scalar-vs-spinor}
\end{figure}

Figure \ref{fig-scalar-vs-spinor} compares the spectra of the massive Klein-Gordon and Dirac fields (studied in \cite{Lasenby-2005-bs}) on the Schwarzschild background. In the limit $M \mu \ll 1$, the spectra follow equation (\ref{hydrogen-spectrum}).  As $M \mu$ is increased, the frequency develops a non-negligible negative imaginary component. At higher couplings, the spin has a significant effect on the frequency levels. In the spin-half case, the $j = l \pm 1/2$ degeneracy is split by the black hole interaction. For couplings $M\mu \gtrsim 0.3$, the (negative) imaginary part of the energy is comparable to the field mass. This means that decay is extremely fast, similar to the Compton time. To put it another way, if $M\mu \gtrsim l$, the state lasts only a few multiples of the light-crossing time for the black hole.

\begin{figure}
\begin{center}
\includegraphics[width=16cm]{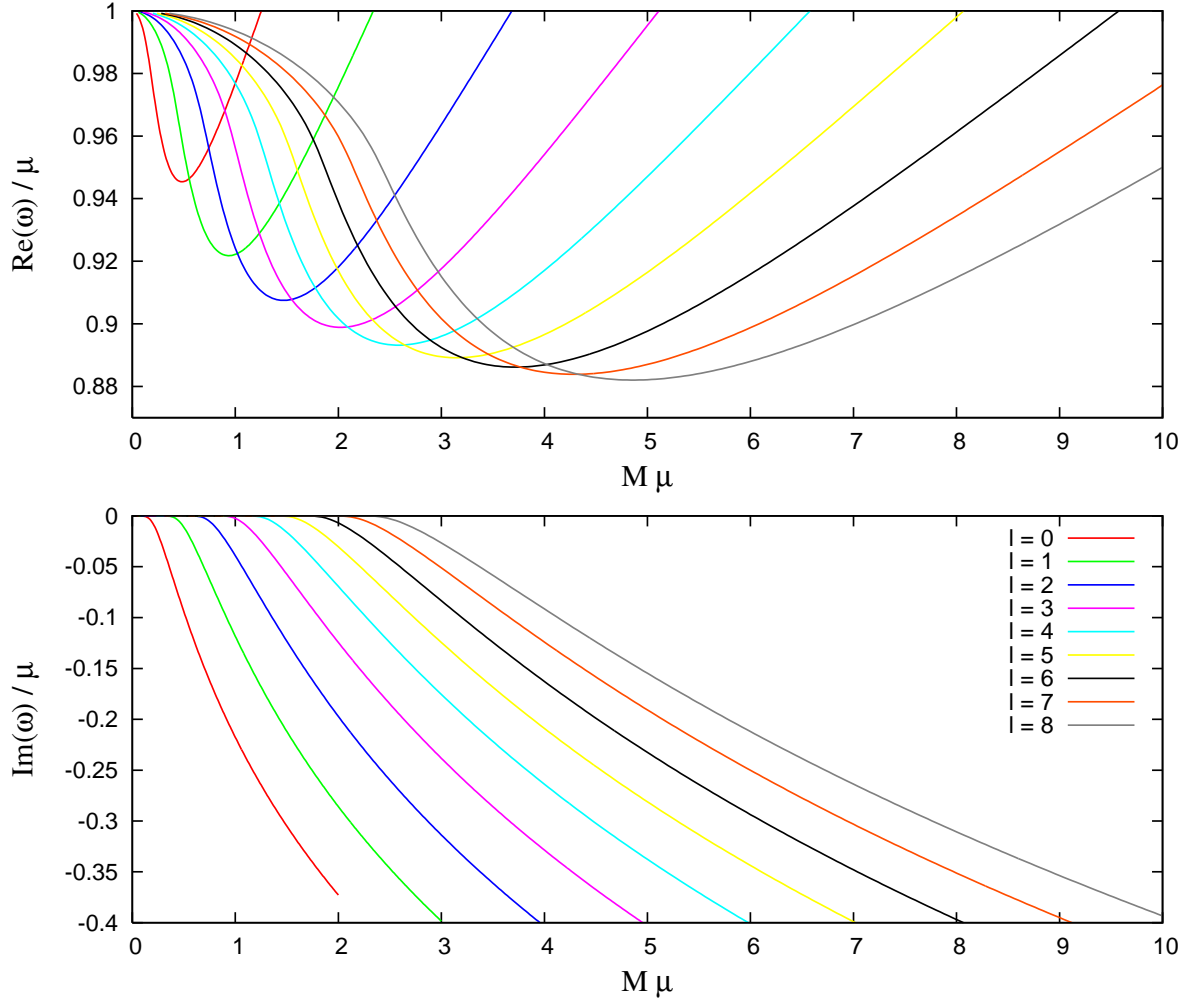}
\end{center}
\caption[]{\emph{The complex frequencies of the lowest-energy Schwarzschild bound states up to $l = 8$.} The top plot shows the oscillation frequency $\text{Re}(\omega / \mu)$, and bottom plot shows the decay rate $\text{Im}(\omega / \mu)$, as a function of the mass coupling $GM\mu / \hbar c$.}
\label{fig-energy-levels}
\end{figure}

Figure \ref{fig-energy-levels} shows the frequency spectrum for scalar states of higher angular momentum, up to $l = 8$. Again, the levels follow the hydrogenic ($1/\overline{n}^2$) spectrum (\ref{hydrogen-spectrum}) in the regime $M \mu \ll l$. At low couplings, the states are quasi-stable. Decay dominates beyond about $M \mu \sim 0.3 l$.  At around $M \mu \sim 0.5 (l+1)$ the real part of the energy reaches a minimum. The maximum `binding energy' offered by this minimum increases with $l$, to around $12 \%$ of the rest mass energy for $l = 8$. It seems unlikely that this energy could be extracted from the black hole, since the state decays very rapidly (with a lifetime similar to the black hole light-crossing time).

\subsection{Bound State Frequencies: Kerr $(a > 0)$\label{sec-bs-kerr}}
\begin{figure}
\begin{center}
\includegraphics[width=13cm]{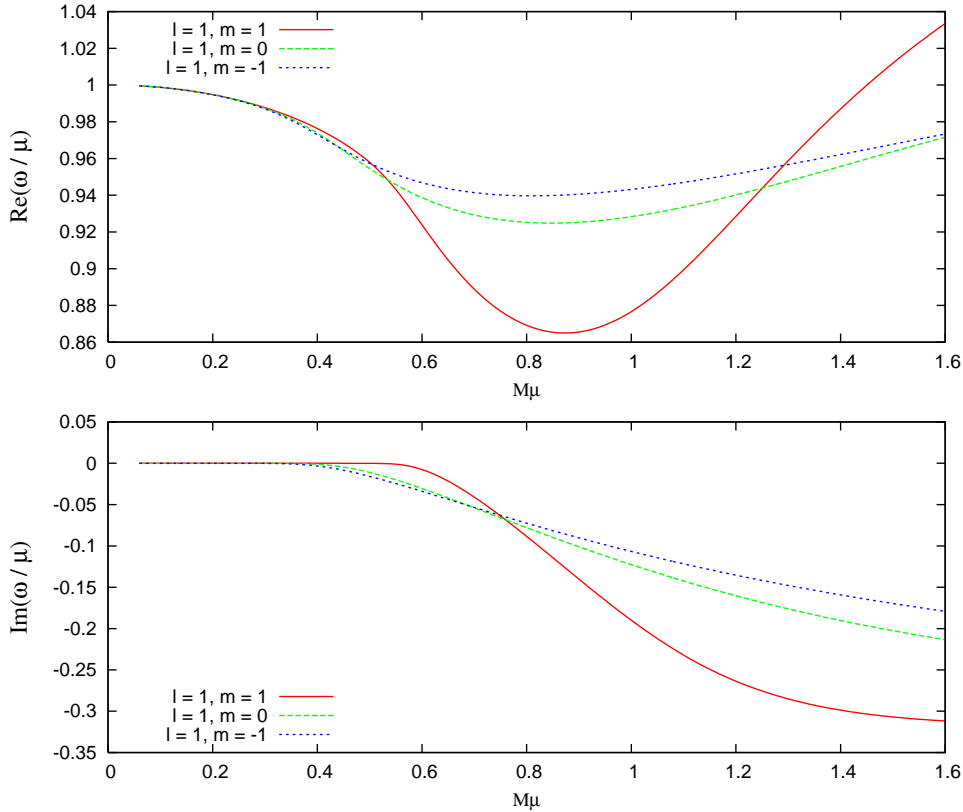}
\end{center}
\caption[]{\emph{Frequency spectrum of the $l =1$,  $m = -1 \ldots 1$ bound states at $a = 0.99$.} The top plot shows the oscillation frequency $\text{Re}(\omega / \mu)$, and bottom plot shows the damping rate $\text{Im}(\omega / \mu)$,, as a function of mass coupling $M\mu$.}
\label{fig-l1m01-1}
\end{figure}

As expected, the introduction of black hole rotation breaks the azimuthal degeneracy. That is, bound states with different azimuthal numbers $m$ occur at different frequencies. Figure \ref{fig-l1m01-1} shows the effect of rotation ($a = 0.99$) on the spectrum of $l = 1$ states. Clearly, the co-rotating state ($m = 1$) behaves very differently to the counter-rotating state ($m = -1$), and at low $M \mu$ its decay is heavily suppressed. At higher couplings, the maximum binding energy is significantly greater for the $m=1$ state than the $m=0$ and $m=-1$ states ($ \sim 13\%$ vs $\sim 7\%$). 

\begin{figure}
\begin{center}
\includegraphics[width=13cm]{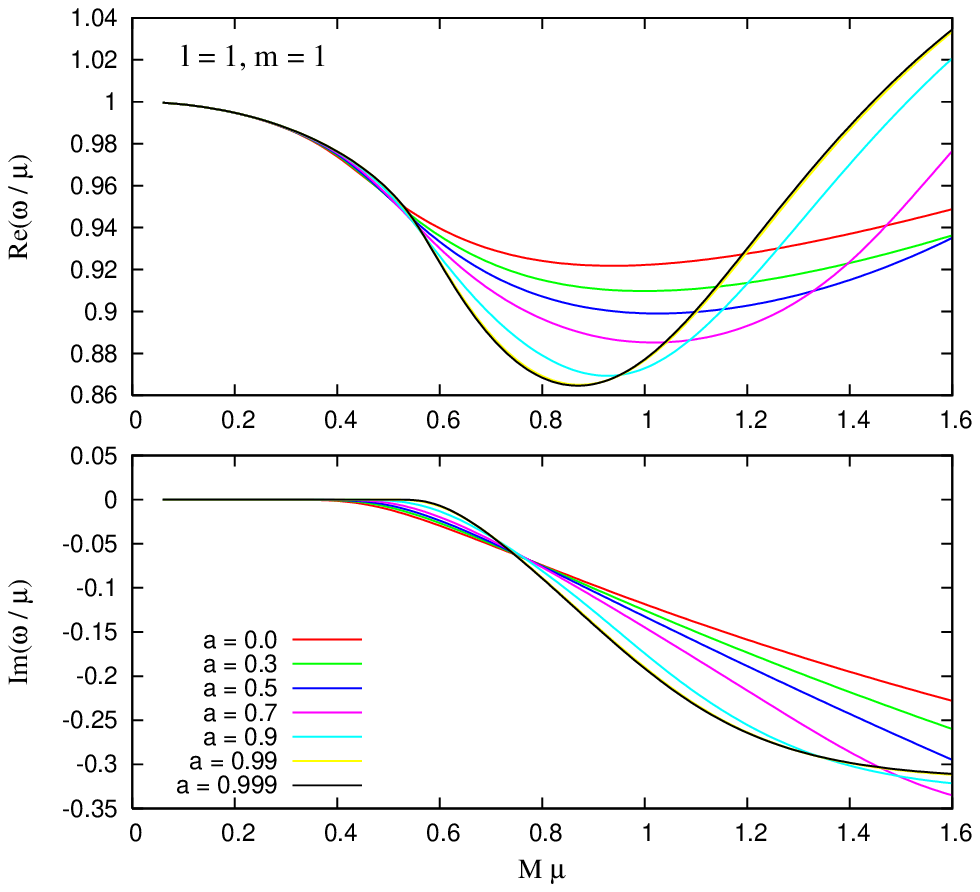}
\end{center}
\caption[]{\emph{Bound state frequencies of the $l =1, m = 1$ state for a range of rotation speeds $a$.} The top plot shows the oscillation frequency $\text{Re}(\omega / \mu)$, and bottom plot shows the damping rate $\text{Im}(\omega / \mu)$, as a function of mass coupling $M\mu$.}
\label{fig-l1m1}
\end{figure}

Maximally co-rotating states (with $m = l$) are of particular interest because of the influence of superradiance. Figure \ref{fig-l1m1} shows the frequency levels of the $l = 1$, $\,m = 1$ state as a function of $M\mu$, for range of rotation speeds. Faster rotation has two effects on the co-rotating spectrum. First, the real energy minimum moves lower, increasing the maximum binding energy from $\sim 8\%$ for $a=0$ to $\sim 13\%$ for $a = 0.99$. Second, faster rotation leads to less damping, at least at low couplings. 

\subsection{The Kerr Instability\label{sec-instability}}

\begin{figure}
\begin{center}
\includegraphics[width=14cm]{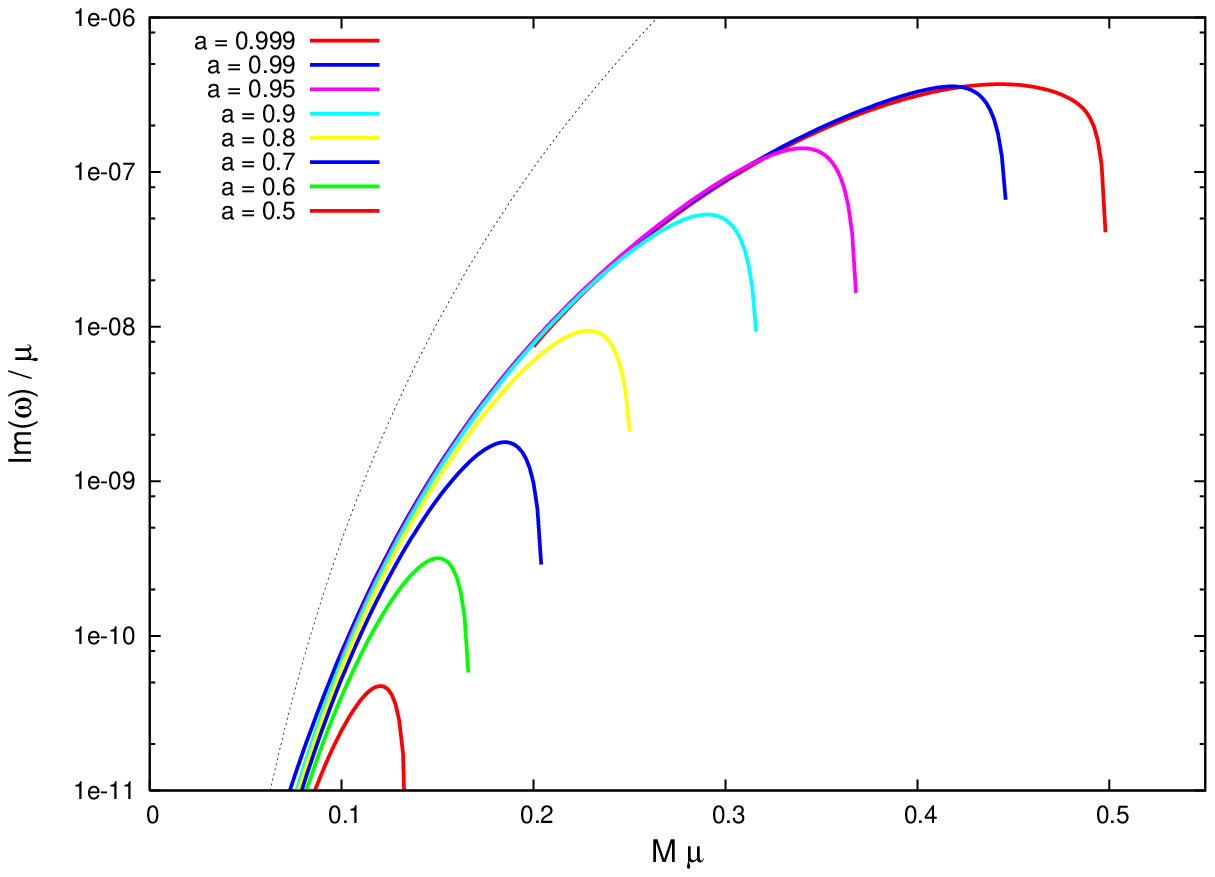}
\end{center}
\caption[]{\small{\emph{Superradiant instability for the $l = 1$, $m = 1$ state.} The growth rate of the $l =1$, $m = 1$ bound state is shown as a function of scalar field mass $\mu$, for a range of black hole rotations $a$. The dotted lines shows Detweiler's approximation \cite{Detweiler-1980}, $\text{Im}(\omega/\mu) \sim (M\mu)^8 / 24$, valid in the limit $M\mu \ll 1$.}}
\label{fig-instability}
\end{figure}

``Zooming in'' on the lower plot of Fig. \ref{fig-l1m1} reveals the Kerr instability: the imaginary part of the frequency is actually positive at low couplings $M\mu$! This effect was predicted by the analysis of section \ref{sec-stress-energy}. Figure \ref{fig-instability} shows the imaginary component $\text{Im}(\omega)$ as a function of coupling $M \mu$. The figure shows that \emph{all} bound states with $\text{Re}(\omega) \lesssim \omega_c$ are unstable. The imaginary component of the frequency reaches a maximum at a coupling just below the superradiant cutoff, $\mu \sim \omega_c$. As expected, faster rotation creates a greater instability. 

\begin{figure}
\begin{center}
\includegraphics[width=14cm]{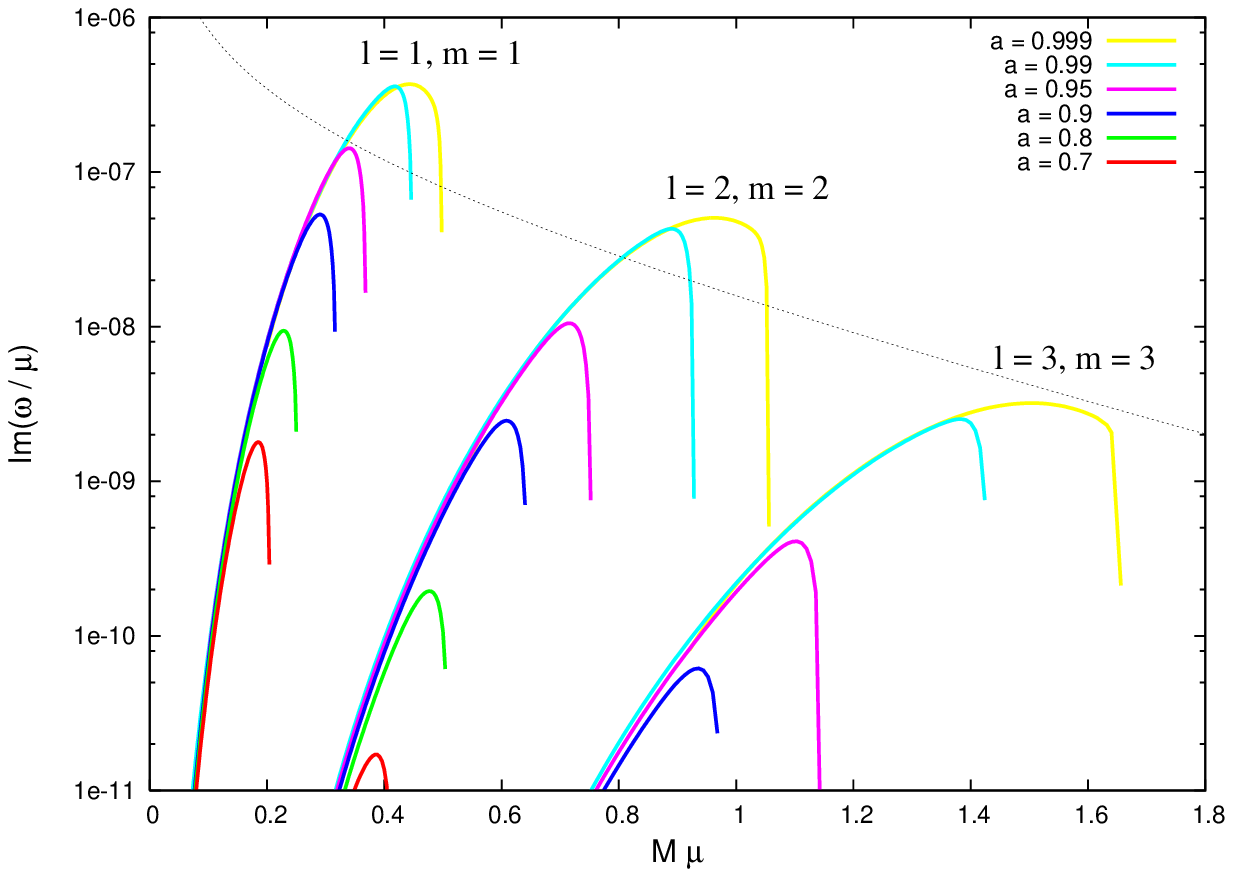}
\end{center}
\caption[]{\emph{Growth rates of the maximally co-rotating modes: $l=m=1$, $l=m=2$ and $l=m=3$.}  The (positive) imaginary part of the frequency is shown as a function of gravitational coupling. The fastest growth occurs for the $l=m=1$ state at $M \mu \approx 0.42$, with $a = 0.999$. The maximum growth rate is approximately $\tau^{-1} = M \text{Im}(\omega) \approx 1.5 \times 10^{-7} (GM/c^3)^{-1}$, where $\tau$ is the e-folding time. The dotted line shows Zouros and Eardley's \cite{Zouros-1979} approximation, $\text{Im}(\omega / \mu) \approx 10^{-7} e^{-1.84 M \mu} / (M \mu)$, valid when $M\mu \gg 1$.}
\label{fig-instability-l2}
\end{figure}

Figure \ref{fig-instability-l2} compares the growth rate of the $l = 1, m = 1$ state with the co-rotating modes of higher angular momentum ($l=2$, $m=2$ and $l=3$, $m=3$). As predicted \cite{Zouros-1979, Detweiler-1980}, the $l =1$, $m= 1$ mode proves to be the most unstable. The dotted line in Fig. \ref{fig-instability-l2} shows the high-coupling approximation (\ref{eq-zouros}).

Table \ref{table-instability} lists some maximum growth rates ($\tau^{-1} = M \, \text{Im}(\omega)$) for the $l = 1$, $m = 1$ mode, for various $a$. Although it is not proved here, we believe that the bound states represent the \emph{most} unstable solutions to the massive wave equations, since no superradiant radiation escapes to infinity. If this is correct, the values in Table \ref{table-instability} represent absolute upper bounds on the growth rate of the scalar field instability.

\begin{table}
\caption{Maximum instability growth rates of the $l = 1$, $m = 1$ state}
\label{table-instability}
\begin{tabular}{c | c c c c c c}
$a$ \brk $0.7$ \brk $0.8$ \brk $0.9$ \brk $0.95$ \brk $0.98$ \brk $0.99$ \; \\
\hline 
$\mu$ \brk $0.187$ \brk $0.231$ \brk $0.293$ \brk $0.343$ \brk $0.393$ \brk $0.421$ \lend
$\tau^{-1}$ \brk $3.33 \times 10^{-10}$ \brk $2.16 \times 10^{-9}$ \brk $1.55 \times 10^{-8}$ \brk $4.88 \times 10^{-8}$ \brk $1.11 \times 10^{-7}$ \brk $1.50 \times 10^{-7}$ 
\end{tabular}
\end{table}

\section{Discussion and Conclusions\label{sec-discussion}}

In this study, we have shown that the instability in the massive scalar field on the Kerr spacetime is greatest when the gravitational coupling is $M \mu \lesssim 0.5$. The $l = 1$, $m = 1$ state is the most unstable, with a maximum growth rate of $\tau^{-1} 1.5 \times \sim 10^{-7} \, (GM/c^3)^{-1}$ at $a = 0.99$. Accurate upper bounds on the growth rate are presented in Table \ref{table-instability}, for a range of $a$. 

Our numerical results are consistent with earlier studies \cite{Zouros-1979, Detweiler-1980, Furuhashi-2004, Cardoso-2005-superr}. For example, Furuhashi and Nambu \cite{Furuhashi-2004} studied the massive scalar instability for the charged, rotating (Kerr-Newman) black hole, using matching methods. For the special case $Q = 0$, $a = 0.98$ they found that the growth rate reaches a maximum value $\tau^{-1} \sim 1.1 \times 10^{-7} \, (GM/c^3)^{-1} $ at $\mu \sim 0.38$ (see Fig. 6 in \cite{Furuhashi-2004}). By comparison, we find $\tau^{-1} \approx 1.11 \times 10^{-7}$ at $\mu \approx 0.39$, for $a = 0.98$.

In recent years, a number of groups \cite{Krivan-1997, Andersson-1998, Andersson-2000, Burko-2004, Strafuss-2005} have studied the massive scalar field in the time domain, using numerical finite-difference codes. For example, Strafuss and Khanna \cite{Strafuss-2005} try perturbing the Kerr hole with a `nearly monochromatic' wave pulse, and study the late-time tail for signs of instability. They report an e-folding time of $\tau \sim 5 \times 10^4$ for the parameter values $a = 0.9999$, $M \mu = 0.25$, $l = 1$, $m = 1$. On the other hand, our results (Fig. \ref{fig-instability}) suggest a minimum e-folding time of around $1.2 \times 10^8$ for these values. Without further details, it is difficult to explain this large discrepancy.

Before concluding, it is worth attempting to assess the observable consequences of the instability, if any. That is, does the massive boson instability actually manifest itself in any physical system? For astrophysical systems, the short answer appears to be: almost certainly not! This is because, for any known massive boson coupled to an astrophysical black hole ($M \gtrsim M_\odot$), the gravitational coupling would be very large, $M \mu \gg 1$. Hence by (\ref{eq-zouros}) the instability is negligible. Nevertheless, this prospect cannot be entirely discounted. If apparently massless bosons turn out to have a small but non-zero rest mass, then the instability could play a role in slowing the rotation of large black holes. For example, for a supermassive black hole ($M \sim 10^{10} M_\odot$), the instability would be significant if there exists a stable boson with mass $\mu \sim 10^{-20} \, \text{eV}$. For comparison, the (lab-based) experimental upper bound on the photon's mass is currently around $\sim 10^{-16} \, \text{eV}$ \cite{Tu-2005}. 

For the superradiant instability to be significant for small `primordial' black holes ($M \lesssim 10^{12}\, \text{kg}$), two conditions must be met \cite{Zouros-1979}. First, the e-folding time should be significantly shorter than the Hawking evaporation lifetime. Second, the instability growth rate should exceed the spontaneous decay rate of the particle.

Let us consider the first condition: is the Kerr instability able to compete with Hawking radiation for small black holes? It is thought that over the age of the universe ($t_{\text{Hubble}} \sim 5 \times 10^{17} \, \text{s}$), primordial black holes smaller than $M \sim 10^{12} \, \text{kg}$ will evaporate through Hawking radiation \cite{Page-1976b}. The light-crossing time for a primordial black hole would be short ($2.5 \times 10^{-24} \, \text{s}$), so the dimensionless growth rate need be no bigger than $\tau^{-1} \sim 10^{-40}$ to be of some significance. Using approximations (\ref{eq-detweiler}) and (\ref{eq-zouros}), this corresponds to a coupling in the range $2.5 \times 10^{-5} \lesssim M \mu \lesssim 50$. For $M \sim 10^{12} \, \text{kg}$ this implies that the instability could arise if the field mass is in the range $10 \, \text{keV} \lesssim M \mu \lesssim 10 \, \text{GeV}$. Repeating this analysis for smaller black holes, $M \sim 10^9\,$kg and $10^6\,$kg, leads to mass ranges $50 \, \text{MeV} \lesssim \mu \lesssim 10^4 \, \text{GeV}$ and $100 \, \text{GeV} \lesssim \mu \lesssim 10^7 \, \text{GeV}$, respectively. These mass ranges span the known particle spectrum. It is therefore highly likely that, at least at some stage during the black hole's evolution, the first condition will be met.

Now we come to the second condition: does instability growth outstrip natural radioactive decay? As a first example, let us consider the neutral pion $\pi^0$. The pion has a mass $\mu \approx 134.96  \,\text{MeV} \approx 2.4 \times 10^{-28} \, \text{kg}$ and a lifetime $\tau_{1/2} \approx 8 \times 10^{-17} \,s$. The instability grows fastest when $M \mu \sim 0.5$, which would correspond to a black hole of mass $M \sim 1 \times 10^{12} \, \text{kg}$. If we take the maximum growth rate of $\tau^{-1} \sim 1.5 \times 10^{-7} GM / c^3$ at $a = 0.99$, this implies a lower bound on the e-folding time of $\tau \sim 1.5 \times 10^{-17}\, \text{s}$. It is a curious coincidence that this the instability growth is of the same order as the rate of radioactive decay! Thus, exponential growth in the neutral pion field is only possible in a very narrow parameter range. Though charged pions $\pi^{\pm}$ are potentially much more stable ($\tau_{1/2} \sim 2.6 \times 10^{-8} \, \text{s}$) than the neutral versions, any charged pair of particles in the vicinity of  the hole would annihilate quickly, thus quenching the instability growth.

A second possible candidate particle is the spin-1 $Z_0$ boson, which mediates the weak force. The $Z_0$ boson is about a thousand times more massive than the pion, but decays $5 \times 10^8$ times faster! Neglecting the effects of spin leads to an estimate of $1.5 \times 10^{-20} \, \text{s}$ for the shortest possible e-folding time. This is far too slow for any instability to arise. Nevertheless, it would be interesting to repeat the frequency-domain analysis for the massive spin-1 field. As is well-known, the superradiance effect is enhanced by the spin of the field \cite{Starobinskii-1973}. For example, the maximum reflection coefficient is $1.003$ for the massless scalar wave, but increases to $1.044$ for electromagnetic waves \cite{Press-1972}. It is natural to suppose that the Kerr instability is similarly enhanced. 

Finally, some recently-proposed theories \cite{Randall-1999b} invoke ``large'' ($l \gg l_p$) extra dimensions to explain why gravity is so much weaker than the other three forces. These theories raise the possibility that higher-dimensional gravitational objects may be created through high energy particle collisions. With $n \ge 1$ additional dimensions, possibilities include black strings (with event horizon topology $S^{1+n} \times R$) and black $p$-branes ($S^{2+n-p} \times R^p$), as well as black holes ($R^{2+n}$) \cite{Kanti-2004}. The stability properties of higher-dimensional black objects have attracted much interest \cite{Gregory-1993, Berti-2003, Cardoso-2005-new}. Black strings (branes) are thought to suffer from the Gregory-Laflamme instability \cite{Gregory-1993} which breaks the black string (brane) into smaller segments. On the other hand, higher-dimensional black holes are not affected by the Gregory-Laflamme instability. A recent study \cite{Cardoso-2005-superr} suggests that higher-dimensional rotating black holes do not suffer a superradiant instability either. Though they still exhibit superradiance, it seems that bound states are prohibited. This conclusion is supported by the apparent absence of stable orbits on higher-dimensional rotating backgrounds \cite{Frolov-2003}. 

\appendix

\section{Non-Relativistic Frequency Spectrum\label{non-rel}}
Here we show that, in the non-relativistic limit, the frequency spectrum of the scalar field bound to a Schwarzschild black hole is given by (\ref{hydrogen-spectrum}). Let us begin by considering the Schwarzschild spacetime described by Painlev\'e-Gullstrand (PG) coordinates \cite{Martel-2001}, 
\begin{equation}
ds^2 = \left(1 - \tMor \right) dt^2 - \sqrt{8M/r} dt dr - dr^2 - d\Omega^2 
\end{equation}
The contravariant metric tensor has components $g^{tt} = 1$, $g^{tr} = -\sqrt{2M/r}$ and $g^{rr} = -(1 - 2M/r)$. It is straightforward to show that the massive Klein-Gordon equation in PG coordinates can be written 
\begin{equation}
\left( \partial_t - \sqrt{\tMor} \, \partial_r \right)^2 \Phi - \frac{3}{2r} \sqrt{\frac{2M}{r}} \left( \partial_t - \sqrt{\tMor} \, \partial_r \right) \Phi - \mathbf{\grad}^2 \Phi + \mu^2 \Phi = 0
\end{equation}
where $\mathbf{\grad}^2$ is the 3D Laplacian operator.

To effect a non-relativistic reduction, we split the field $\Phi$ into two components $\chi_1$ and $\chi_2$, defined by
\begin{align}
\chi_1 &= \tfrac{1}{2} \left( \Phi + \frac{i}{\mu} \left( \partial_t - \sqrt{\tMor} \, \partial_r \right) \Phi \right) , \\
\chi_2 &= \tfrac{1}{2} \left( \Phi - \frac{i}{\mu} \left( \partial_t - \sqrt{\tMor} \, \partial_r \right) \Phi \right) ,
\end{align}
so that
\begin{equation}
\chi_1 + \chi_2 = \Phi \quad \quad \text{ and } \quad \quad \chi_1 - \chi_2 = \frac{i}{\mu} \left( \partial_t - \sqrt{\tMor} \, \partial_r \right) \Phi .
\end{equation}
This decomposition leads to the pair of coupled equations,
\begin{align}
\left( i\partial_t - \mu \right) \chi_1  &=  -\frac{1}{2 \mu} \grad^2 (\chi_1 + \chi_2) + i \sqrt{\tMor} \, \partial_r \chi_1 + \frac{3i}{4r} \sqrt{\tMor} \, (\chi_1 - \chi_2) , \\
\left( i\partial_t + \mu \right) \chi_2  &=  +\frac{1}{2 \mu} \grad^2 (\chi_1 + \chi_2) + i \sqrt{\tMor} \, \partial_r \chi_2 + \frac{3i}{4r} \sqrt{\tMor} \, (\chi_2 - \chi_1) .
\end{align}
In the non-relativistic limit, we make the assumption that $\omega \sim \mu $ and the approximation $\chi_2 \ll \chi_1$. Equally well, we could make the assumption that $\omega \sim -\mu$ to recover the non-relativistic antiparticle spectrum. This assumption leads to the Schr\"odinger equation
\begin{equation}
E_{NR} \chi_1  = -\frac{1}{2 \mu} \grad^2 \chi_1 + i \sqrt{\frac{2M}{r}} \left( \partial_r  + \frac{3}{4r} \right) \chi_1 \label{schro1}
\end{equation}
where $E_{NR} = \omega - \mu$. With a simple substitution, $\chi_1 = \psi \exp(i \mu \sqrt{8Mr})$, equation (\ref{schro1}) can be transformed to the familiar form
\begin{equation}
E_{NR} \psi = -\frac{1}{2\mu} \grad^2 \psi - \frac{M \mu}{r} \psi . \label{schro2}
\end{equation}
This is the hydrogenic Schr\"odinger equation, but with the fine-structure constant $\alpha_{EM} = e^2 / 4\pi\epsilon_0 \hbar c$ replaced by the gravitational coupling $\alpha_G = G M \mu / \hbar c$. Hence the non-relativistic wavefunctions are hydrogenic, and the energy levels are given by (\ref{hydrogen-spectrum}).

\section{Quasinormal Mode Frequencies\label{sec-tables}}
\begin{table}[!ht]
\caption{Quasinormal mode frequencies for $l = 1, m = -1$\label{table-qnm1}}
\begin{tabular}{| l | l l | l l | l l | l l |}
\hline 
  \brk $\mu = 0.0$ \brk \brk $\mu = 0.1$ \brk \brk $\mu = 0.2$ \brk \brk $\mu = 0.3$ \brk \lend 
 a \brk $\text{Re}(\omega)$ \brk $-\text{Im}(\omega)$ \brk $\text{Re}(\omega)$ \brk $-\text{Im}(\omega)$ \brk $\text{Re}(\omega)$ \brk $-\text{Im}(\omega)$ \brk $\text{Re}(\omega)$ \brk $-\text{Im}(\omega)$ \; \\
\hline
\, 0.0 \brk	0.292936 \brk  0.097660 \brk 	0.297416 \brk  0.094957 \brk  0.310957 \brk  0.086593 \brk  0.333777 \brk   0.071658 \lend
 0.1 \brk	0.285570 \brk  0.097626 \brk	0.290234 \brk  0.094747 \brk	0.304341 \brk  0.085845 \brk	0.328135 \brk   0.069968 \lend
 0.2 \brk	0.278833 \brk  0.097475 \brk  0.283672 \brk  0.094427 \brk  0.298318 \brk  0.085009 \brk  0.323048 \brk   0.068231 \lend
 0.3 \brk	0.272635 \brk  0.097228 \brk  0.277641 \brk  0.094019 \brk  0.292803 \brk  0.084108 \brk  0.318436 \brk   0.066465 \lend
 0.4 \brk	0.266901 \brk  0.096901 \brk  0.272068 \brk  0.093537 \brk 	0.287726 \brk  0.083154 \brk	0.314235 \brk   0.064680 \lend
 0.5 \brk	0.261572 \brk  0.096505 \brk	0.266893 \brk  0.092994 \brk	0.283032 \brk  0.082158 \brk	0.310392 \brk   0.062887 \lend
 0.6 \brk	0.256596 \brk  0.096051 \brk	0.262066 \brk  0.092399 \brk	0.278671 \brk  0.081130 \brk	0.306865 \brk   0.061091 \lend
 0.7 \brk	0.251928 \brk  0.095547 \brk	0.257544 \brk  0.091760 \brk	0.274604 \brk  0.080076 \brk	0.303615 \brk   0.059298 \lend
 0.8 \brk	0.247531 \brk  0.095000 \brk	0.253289 \brk  0.091085 \brk	0.270795 \brk  0.079004 \brk	0.300613 \brk   0.057513 \lend
 0.9 \brk	0.243371 \brk  0.094422 \brk	0.249270 \brk  0.090383 \brk	0.267217 \brk  0.077919 \brk	0.297831 \brk   0.055740 \lend
 0.95 \brk	0.241372 \brk  0.094124 \brk	0.247341 \brk  0.090025 \brk	0.265506 \brk  0.077375 \brk	0.296516 \brk   0.054859 \lend
 0.99 \brk	0.239810 \brk  0.093882 \brk	0.245834 \brk  0.089736 \brk	0.264173 \brk  0.076939 \brk  0.295520 \brk     0.054158 \lend
 0.995 \brk	0.239616 \brk   0.093852 \brk	0.245647 \brk  0.089700 \brk	0.264006 \brk  0.076884 \brk	0.295290 \brk   0.054039 \\
\hline
\end{tabular}
\end{table}

\begin{table}[!ht]
\caption{Quasinormal mode frequencies for $l = 2, m = 2$\label{table-qnm2}}
\begin{tabular}{| l | l l | l l | l l | l l |}
\hline 
  \brk $\mu = 0.0$ \brk \brk $\mu = 0.1$ \brk \brk $\mu = 0.2$ \brk \brk $\mu = 0.3$ \brk \lend 
 a \brk $\text{Re}(\omega)$ \brk $-\text{Im}(\omega)$ \brk $\text{Re}(\omega)$ \brk $-\text{Im}(\omega)$ \brk $\text{Re}(\omega)$ \brk $-\text{Im}(\omega)$ \brk $\text{Re}(\omega)$ \brk $-\text{Im}(\omega)$ \; \\
\hline 
\,0.0 \brk	0.483644 \brk  0.096759 \brk  0.486804 \brk  0.095675 \brk  0.496327 \brk  0.092389 \brk  0.512346 \brk  0.086795 \lend 
0.1 \brk	0.499482 \brk  0.096666 \brk	0.502456 \brk  0.095674 \brk	0.511419 \brk  0.092663 \brk	0.526497 \brk  0.087528 \lend
0.2 \brk	0.517121 \brk  0.096382 \brk	0.519901 \brk  0.095483 \brk	0.528281 \brk  0.092755 \brk  0.542378 \brk  0.088092 \lend
0.3 \brk	0.536979 \brk  0.095839 \brk	0.539557 \brk  0.095036 \brk	0.547326 \brk  0.092595 \brk	0.560397 \brk  0.088418 \lend
0.4 \brk	0.559647 \brk  0.094931 \brk	0.562011 \brk  0.094226 \brk	0.569137 \brk  0.092082 \brk	0.581127 \brk  0.088406 \lend
0.5 \brk	0.585990 \brk  0.093494 \brk	0.588127 \brk  0.092890 \brk	0.594568 \brk  0.091052 \brk	0.605408 \brk  0.087895 \lend
0.6 \brk	0.617364 \brk  0.091245 \brk	0.619256 \brk  0.090746 \brk	0.624959 \brk  0.089224 \brk	0.634559 \brk  0.086607 \lend
0.7 \brk	0.656099 \brk  0.087649 \brk	0.657722 \brk  0.087259 \brk	0.662614 \brk  0.086068 \brk	0.670848 \brk  0.084018 \lend
0.8 \brk	0.706823 \brk  0.081520 \brk	0.708138 \brk  0.081245 \brk	0.712102 \brk  0.080407 \brk 	0.718777 \brk  0.078961 \lend
0.9 \brk	0.781638 \brk  0.069289 \brk	0.782570 \brk  0.069142 \brk	0.785380 \brk  0.068690 \brk	0.790113 \brk  0.067911 \lend
0.95 \brk	0.840982 \brk  0.056471 \brk	0.841653 \brk  0.056395 \brk	0.843677 \brk  0.056163 \brk	0.847088 \brk  0.055762 \lend
0.99 \brk	0.928028 \brk  0.031063 \brk	0.928353 \brk  0.031054 \brk	0.929336 \brk  0.031026 \brk	0.930994 \brk  0.030977 \lend
0.995 \brk	0.949522 \brk  0.023104 \brk 	0.949762 \brk  0.023104 \brk	0.950487 \brk  0.023102 \brk	0.951712 \brk   0.023100 \\
\hline
\end{tabular}
\end{table}

\begin{acknowledgments}
SD would like to thank Calvin Smith, Vitor Cardoso, Elizabeth Winstanley and Marc Casals for helpful discussions and proof-reading, and the Astrophysics Group at Cambridge University for the continued use of their computing facilities.  
\end{acknowledgments}

\bibliographystyle{acm}

\end{document}